\documentclass[aps,prc,twocolumn,10pt,superscriptaddress]{revtex4-2}
\usepackage{amsmath,amsfonts,amssymb}
\usepackage{graphicx}
\usepackage[normalem]{ulem}
\usepackage{slashed} 
\usepackage[dvipsnames]{xcolor}

\usepackage{hyperref}
\hypersetup{
    colorlinks,
    unicode=True,
    linkcolor=Blue,
    citecolor=Blue,
    urlcolor=Blue,
    pdfauthor={Gaia Ingrosso},
}

\newcommand{\ubar}{\bar{u}}

\newcommand{\dpi}{(2\pi)}
\newcommand{\VX}[3]{V^{#1,#2}_{#3}}
\newcommand{\QP}[1]{\Lambda(#1)}
\newcommand{\GV}[3]{\Omega^{#1,#2}(#3)}

\newcommand{\Msq}[1]{|\bar{\mathcal{M}}_{#1}|^2}
\newcommand{\GL}[3]{\varepsilon^{#1}_{#2}(#3)}
\newcommand{\GP}[3]{\Pi_{#1#2}(#3)}

\newcommand{\GI}[1]{{\color{black}#1}} 
\newcommand{\EB}[1]{{\color{black}#1}} 

\begin{document}

\title{Transport coefficients of strongly interacting quark-gluon plasma including elastic and inelastic scattering within the dynamical quasiparticle model}

\author{Gaia Ingrosso}
\email{ingrosso@itp.uni-frankfurt.de}
\affiliation{Institut f\"ur Theoretische Physik, Johann Wolfgang Goethe	University, Max-von-Laue-Str. 1, 60438 Frankfurt, Germany}

\author{Olga Soloveva}
 \affiliation{GSI Helmholtzzentrum f\"ur Schwerionenforschung GmbH, Planckstraße 1, 64291 Darmstadt, Germany}
 \affiliation{Institut f\"ur Theoretische Physik, Johann Wolfgang Goethe	University, Max-von-Laue-Str. 1, 60438 Frankfurt, Germany}
 \affiliation{Helmholtz Research Academy Hessen for FAIR (HFHF), GSI Helmholtz	Center for Heavy Ion Physics, Campus Frankfurt, 60438 Frankfurt, Germany}
 
\author{Ilia Grishmanovskii}
\affiliation{Institut f\"ur Theoretische Physik, Johann Wolfgang Goethe	University, Max-von-Laue-Str. 1, 60438 Frankfurt, Germany}

\author{Taesoo Song}
\affiliation{GSI Helmholtzzentrum f\"ur Schwerionenforschung GmbH, Planckstraße 1, 64291 Darmstadt, Germany}

\author{Elena Bratkovskaya}
\affiliation{GSI Helmholtzzentrum f\"ur Schwerionenforschung GmbH, Planckstraße 1, 64291 Darmstadt, Germany}
\affiliation{Institut f\"ur Theoretische Physik, Johann Wolfgang Goethe	University, Max-von-Laue-Str. 1, 60438 Frankfurt, Germany}
\affiliation{Helmholtz Research Academy Hessen for FAIR (HFHF), GSI Helmholtz	Center for Heavy Ion Physics, Campus Frankfurt, 60438 Frankfurt, Germany}

\date{\today}
\begin{abstract}
We study the impact of inelastic gluon-radiation \GI{and absorption} processes on the transport coefficients of the quark-gluon plasma within the dynamical quasiparticle model (DQPM) in the temperature--baryon-chemical-potential plane $(T,\mu_B)$. Extending the \GI{$2\rightarrow2$} baseline established in previous DQPM calculations, we include \GI{gluon-radiation $2\rightarrow3$ and the inverse gluon-absorption $3\rightarrow2$} scattering channels with massive partons and effective DQPM propagators and vertices. The corresponding momentum-dependent interaction rates and relaxation times are used to calculate the shear viscosity, bulk viscosity, electric conductivity, and baryon diffusion coefficient as functions of temperature $T$ and baryon chemical potential $\mu_B$.
 \GI{Within the relaxation time approximation, we find that $2\leftrightarrow3$ contributions} systematically reduce all considered transport coefficients relative to the \GI{$2\rightarrow2$} only results, in accordance with the decrease of the relaxation times. In the thermal regime explored here, however, this reduction remains moderate, since the investigated inelastic rates stay below the \GI{$2\rightarrow2$} ones over the considered $(T,\mu_B)$ range. The inelastic channels become more relevant mainly for partonic scatterings at large momenta, which are thermally suppressed in the strongly interacting QGP. 
At $\mu_B=0$, the resulting $\eta/s$, $\zeta/s$, and $\sigma_Q/T$ are compatible with available lattice-QCD estimates within uncertainties. At finite $\mu_B$, our results provide predictions for the transport properties of QCD matter relevant for beam-energy-scan programs.
\end{abstract}

\maketitle

\section{Introduction}
\label{sec:intro}

Relativistic heavy-ion collisions provide a unique experimental tool to create and study strongly interacting QCD matter under extreme conditions. By varying the collision energy, present and future programs at RHIC-BES \cite{Bzdak:2019pkr}, FAIR (CBM) \cite{CBM:2016kpk}, and NICA (MPD/BM@N) \cite{Golovatyuk:2016} explore the properties of the quark-gluon plasma (QGP) from nearly baryon-symmetric matter at $\mu_B \simeq 0$ to baryon-rich matter at sizable baryon chemical potential.

Transport coefficients are key quantities for characterizing the QGP beyond its equilibrium equation of state. They determine the response of the medium to gradients and external perturbations and thus control the space-time evolution of matter created in heavy-ion collisions. The shear viscosity $\eta$ governs momentum diffusion and the damping of anisotropic flow, while the bulk viscosity $\zeta$ reflects the breaking of conformal symmetry and is expected to become important near the transition region. The electric conductivity $\sigma_Q$ determines the response of the QGP to electromagnetic fields, whereas at finite baryon density the baryon diffusion coefficient $\kappa_B$ controls the transport of net-baryon number.

The determination of transport coefficients from QCD remains challenging. Lattice-QCD calculations provide reliable constraints on equilibrium thermodynamics at $\mu_B=0$, where the transition from hadronic matter to the QGP is a smooth crossover, but the extraction of real-time transport coefficients from Euclidean correlators is difficult. At finite $\mu_B$, the sign problem further limits first-principles calculations, especially in the density range relevant for intermediate-energy heavy-ion collisions. Therefore, effective models and microscopic transport approaches are essential for exploring the transport properties of QCD matter at finite temperature and baryon chemical potential.

A variety of approaches has been developed for this purpose, including quasiparticle models, NJL- and PNJL-type models, chiral effective models, holographic approaches, and microscopic transport calculations. Although many of these approaches can be constrained to reproduce similar lattice-QCD equations of state at $\mu_B=0$, their predictions for transport coefficients may differ substantially, reflecting different assumptions about quasiparticle properties, interaction rates, and critical dynamics
\cite{Sasaki:2008fg,Bluhm:2010qf,Marty:2013ita,Zhao:2020zly,Bandyopadhyay:2023lqf,Singh:2025nri}. 

Recent studies have addressed the temperature and chemical-potential dependence of viscosities, electric conductivity, baryon diffusion, and the diffusion matrix of conserved charges within kinetic theory, quasiparticle models, Kubo-based approaches, and hadronic transport
\cite{Madni:2024visc,Madni:2024cond,Greif:2017byw,Fotakis:2019nbq,Fotakis:2021diq,Das:2021bkz,Dey:2024kubo,Ohanaka:2026gla,Parisi:2025gwq,Parisi:2026ufk}. 
In parallel, Bayesian analyses of beam-energy-scan data within viscous hydrodynamic and hybrid frameworks have started to constrain the temperature and baryon-density dependence of QGP transport coefficients \cite{Shen:2023bes,Jahan:2024bes}.

A particularly suitable framework for studying the strongly interacting QGP is the dynamical quasiparticle model (DQPM)
\cite{Peshier:2005pp,Cassing:2008nn}. In the DQPM, the effective degrees of freedom are massive quarks, antiquarks, and gluons with broad spectral functions. The real parts of the partonic self-energies define dynamically generated quasiparticle masses, while the imaginary parts are related to spectral widths. The model parameters are fixed by requiring consistency with lattice-QCD thermodynamics at $\mu_B=0$, while extensions to finite $\mu_B$ allow one to explore the baryon-rich region relevant for beam-energy-scan experiments.

Within this framework, the DQPM has been successfully applied to the description of equilibrium QGP properties and to the calculation of transport coefficients such as the specific shear viscosity, the bulk viscosity, and the diffusion coefficients of conserved charges
\cite{Moreau:2019vhw,Soloveva:2019xph,Fotakis:2021diq,Soloveva:2023tvj,Grishmanovskii:2024qtm,Grishmanovskii:2025mnc}. These quantities also provide microscopic input for the Parton-Hadron-String Dynamics (PHSD) transport approach \cite{Cassing:2009vt,Cassing:2008nn,Linnyk:2015rco,Moreau:2019vhw}. Previous DQPM studies at finite baryon chemical potential have shown that transport coefficients exhibit a nontrivial dependence on both $T$ and $\mu_B$ and may be sensitive to the structure of the QCD phase diagram. In particular, the DQPM with a critical point, DQPM-CP, has been used to investigate the possible impact of a CEP and a first-order phase transition on transport properties \cite{Soloveva:2021quj,Soloveva:2020hpr}.

In Ref.~\cite{Grishmanovskii:2023gog}, the DQPM was extended to include thermal gluon-radiation processes in massive partonic scatterings. In particular, the inelastic reactions $q+q \rightarrow q+q+g$ and $q+g \rightarrow q+g+g$, as well as the corresponding channels involving antiquarks, were evaluated by calculating explicitly the leading-order Feynman diagrams for $2\rightarrow3$ processes with effective DQPM propagators and vertices. This provided, for the first time within the DQPM, a microscopic treatment of radiative partonic processes without additional approximations or simplifications.
In that study, the momentum averaged interaction rates $\Gamma$ and the corresponding relaxation times $\tau$ of quarks and gluons in the strongly interacting QGP were evaluated as functions of temperature at $\mu_B=0$. It was found that $2\rightarrow2$ scatterings give the dominant contribution to the total interaction rate, whereas the inelastic $2\rightarrow3$ transition rates are strongly suppressed. Consequently, the relaxation times obtained by including both elastic and inelastic processes are only slightly shorter than those calculated from elastic scatterings alone. This indicates that radiative processes with the emission of massive gluons are substantially suppressed in the non-perturbative QGP medium described by the DQPM.

In the present work, we investigate the impact of inelastic \GI{gluon-emission and absorption} processes on the transport coefficients of the strongly interacting QGP. Specifically, we calculate the shear and bulk viscosities, the electric conductivity, and the baryon diffusion coefficient within the DQPM as functions of temperature $T$ and baryon chemical potential $\mu_B$. By comparing results obtained from $2\rightarrow2$ scatterings only with those including both $2\rightarrow 2$ and \GI{$2\leftrightarrow 3$} channels, we quantify the relative importance of such inelastic partonic reactions for the transport properties of the QGP. 
Furthermore, a baseline for these transport coefficients that also accounts for inelastic $2 \leftrightarrow 3$ scattering allows one to quantify the role of gluon-emission and absorption processes in microscopic inputs for dynamical modeling of the QGP, including recent studies of early-time and out-of-equilibrium spacetime evolution \cite{Barata:2025zku}.

Following our earlier studies \cite{Moreau:2019vhw,Soloveva:2019xph,Fotakis:2021diq,
Soloveva:2023tvj,Grishmanovskii:2024qtm,Grishmanovskii:2025mnc}, the transport coefficients are evaluated from microscopic collision rates determined by the effective coupling and partonic propagators of the DQPM.
In this way, the relaxation times entering the RTA are directly related to the underlying partonic interaction dynamics. 
The role of inelastic processes has recently been investigated in Ref.~\cite{Grishmanovskii:2025mnc} for the spatial diffusion coefficient of heavy quarks propagating in the strongly interacting QGP. It was shown that radiative reactions provide only a minor correction to the heavy-quark diffusion coefficient, since the latter is dominated by the low-momentum region, where inelastic channels are suppressed relative to elastic scatterings.
In the present work, we extend this analysis to the transport coefficients of the bulk QGP medium. We focus on the temperature $T$ and baryon-chemical-potential $\mu_B$ dependence of the shear and bulk viscosities, the electric conductivity, and the baryon diffusion coefficient. The results provide microscopic input for hydrodynamic and transport simulations of heavy-ion collisions at finite net-baryon density and contribute to constraining the transport properties of QCD matter in the region explored by present and future heavy-ion experiments.

The paper is organized as follows: we start by recalling the basis of the DQPM model in Sec. \ref{sec:DQPM}. Then, we present our results for the transport coefficients including $2\rightarrow2$ and $2\leftrightarrow3$ reactions in Sec. \ref{sec:TC}.
We finish the paper with a summary in Sec. \ref{sec:summary}. 

\section{DQPM}
\label{sec:DQPM}

We start with a brief reminder of the basis of the DQPM and refer the reader to Ref.~\cite{Soloveva:2020hpr} for a more detailed description of the latest version of the DQPM.

The Dynamical Quasiparticle Model (DQPM) \cite{Peshier:2005pp,Cassing:2007nb,Cassing:2007yg,Berrehrah:2016vzw,Moreau:2019vhw,Soloveva:2020hpr} is an effective model that describes the QGP in terms of strongly interacting quarks and gluons. This model is based on fitting the properties of these particles in order to reproduce the results of lattice QCD calculations in thermal equilibrium and at vanishing chemical potential. 

\noindent
The quasiparticles in the DQPM are characterized by the following properties: \\
$\bullet$ {\it Dressed} propagators, i.e., single-particle (two-point) Green's functions, have the form
\begin{equation}
  G^{R}_j (\omega, {\bf p}) = \frac{1}{\omega^2 - {\bf p}^2 - m_j^2 + 2 i \gamma_j \omega}
\label{eq:propdqpm}
\end{equation}
for quarks, antiquarks, and gluons ($j = q,\bar q,g$), using $\omega=p_0$ for energy, the widths $\gamma_{j}$ and the masses $m_{j}$. \\
$\bullet$ The model uses complex self-energies for gluons, $\Pi = m_g^2-2i \omega \gamma_g$, and for (anti)quarks, $\Sigma_{q} = m_{q}^2 - 2 i \omega \gamma_{q}$, where the real part of the self-energies is associated with dynamically generated thermal masses, while the imaginary part provides information about the lifetime and reaction rates of the particles.

The spectral functions in the DQPM are no longer $\delta$ functions, but have a finite width $\gamma_{j}$ \cite{Linnyk:2015rco}:
\begin{align}
\label{eq:spectral_function}
  \rho_{j}(\omega,{\bf p}) &= \frac{\gamma_{j}}{\tilde{E}_j}
  \left(\frac{1}{(\omega-\tilde{E}_j)^2+\gamma^{2}_{j}}
  -\frac{1}{(\omega+\tilde{E}_j)^2+\gamma^{2}_{j}}\right) 
  \nonumber\\
  &\equiv \frac{4\omega\gamma_j}{\left( \omega^2 - {\bf p}^2 - m^2_j \right)^2 + 4\gamma^2_j \omega^2}
\end{align}
Here, $\tilde{E}_{j}^2({\bf p})={\bf p}^2+m_{j}^{2}-\gamma_{j}^{2}$. The spectral function is antisymmetric in $\omega$ and normalized as
\begin{equation}
\label{eq:spectral_function_norm}
  \int\limits_{-\infty}^{\infty}\frac{d\omega}{2\pi}\
  \omega \ \rho_{j}(\omega,{\bf p})=
  \int\limits_{0}^{\infty} d\omega \frac{\omega}{\pi}\ 
  \rho_{j}(\omega,{\bf p})=1.
\end{equation}

$\bullet$ A model ansatz is used for the masses $m_{j}(T,\mu_q)$ and widths $\gamma_{j}(T,\mu_q)$ as functions of the temperature $T$ and the quark chemical potential $\mu_q$.

With the quasiparticle properties (or propagators) fixed as described above, one can evaluate thermodynamic quantities such as the entropy density $s(T,\mu_B)$, the pressure $P(T,\mu_B)$, and energy density $\epsilon(T,\mu_B)$ in a straightforward manner by starting with the entropy density and number density in the propagator representation from Baym \cite{Vanderheyden:1998ph,Blaizot:2000fc}.

By comparison of the entropy density -- computed within the DQPM framework -- to the lQCD data, one can fix the few parameters used in the ansatz for quasiparticle masses and widths.


\subsection{Quasiparticle properties}

The quasiparticle pole masses for gluons and quarks are
defined, inspired  by the asymptotic HTL masses
\cite{Bellac:2011kqa,Linnyk:2015rco}, by
	\begin{equation}
	 m^2_{g}(T,\mu_\mathrm{B})=C_g \frac{g^2(T,\mu_\mathrm{B})}{6}T^2\left(1+\frac{N_{f}}{2N_{c}}
	+\frac{1}{2}\frac{\sum_{q}\mu^{2}_{q}}{T^2\pi^2}\right)
		  \label{polemass_g_dqpm}
	\end{equation}
	
	\begin{equation}
	  m^2_{q(\bar q)}(T,\mu_\mathrm{B})=C_q \frac{g^2(T,\mu_\mathrm{B})}{4}T^2\left(1+ \frac{\mu^{2}_{q}}{T^2\pi^2}\right)
	  \label{polemass_q_dqpm},
	\end{equation}
	where $N_{c}=3$ and $N_{f}=3$ denote the number of colors and the number of flavors, respectively. $C_q = \frac{N_c^2 - 1}{2 N_c} = 4/3$ and $C_g = N_c = 3$ are the QCD color factors for quarks and for gluons, respectively. The strange quark has a larger bare mass which needs to be considered in its dynamical quasiparticle pole mass. We fix $m_s(T,\mu_\mathrm{B})= m_{u}(T,\mu_\mathrm{B})+ \Delta m$ and $\Delta m \approx$ 30 MeV \cite{Moreau:2019vhw}.	

	Furthermore, the quasiparticles in DQPM have thermal widths, which are adopted in the form 
	\begin{equation}
		\gamma_{j}(T,\mu_\mathrm{B}) = \frac{1}{3} C_j \frac{g^2(T,\mu_\mathrm{B})T}{8\pi}\ln\left(\frac{2c_m}{g^2(T,\mu_\mathrm{B})}+1\right).
	\label{eq:widths}
	\end{equation}
The parameter $c_m$, which is related to a magnetic cut-off, is fixed to $c_m = 14.4$.

In the DQPM, the value of $g^2$ is extracted from lQCD by utilizing a parametrization method introduced in Ref.~\cite{Berrehrah:2015vhe}, where it has been shown that for a given value of $g^2$, the ratio $s(T,g^2)/T^3$ is almost constant for different temperatures, i.e., ${\frac{\partial}{\partial T}} (s(T,g^2)/T^3)=0$. Therefore, the entropy density $s$ and the dimensionless equation of state in the DQPM are functions of the effective coupling only, i.e., $s(T,g^2)/s_{SB}(T) = f(g^2)$, where $s^{QCD}_{SB} = 19/9 \pi^2T^3$ is the Stefan-Boltzmann entropy density. Thus, by inverting the $f(g^2)$ function, the coupling constant $g^2$ can be directly obtained from the parametrization of lQCD data for the entropy density $s(T,\mu_B=0)$ at zero baryon chemical potential:
\begin{equation}
\label{eq:g2mub0}
  g^2(T,\mu_B=0) = d \left( (s(T,0)/s^{QCD}_{SB})^e - 1 \right)^f.
\end{equation}
Here $d = 169.934, e = -0.178434$, and $f = 1.14631$ are the dimensionless parameters obtained by adjusting the quasiparticle entropy density $s(T,\mu_B=0)$ to the lQCD data provided by the BMW Collaboration \cite{Borsanyi:2012cr,Borsanyi:2013bia}.

The effective coupling at finite baryon chemical potential $\mu_\mathrm{B}$ is obtained by applying the 'scaling hypothesis' introduced in \cite{Cassing:2007nb}. It assumes that $g^2$ is a function of the ratio of the effective temperature
	 \begin{equation}
	 T^* = \sqrt{T^2+\mu^2_q/\pi^2}
	 \label{eq:tstar}
	 \end{equation}
 (where the quark chemical potential is defined as $\mu_q=\mu_u=\mu_s=\mu_B/3$ ) and the $\mu_\mathrm{B}$-dependent critical temperature $T_C(\mu_\mathrm{B})$ as 
	 	\begin{equation}
	    T_C(\mu_B) = T_C(0) (1- a\mu_B^2)^{1/2},
	    \label{eq:dqpmTc}
	\end{equation}
where $T_C(0)$ is the critical temperature at vanishing chemical potential ($\approx 0.158$ GeV) and $a = 0.974$ GeV$^{-2}$.
\EB{
 We recall that the expression of $T_c(\mu_q)$ in Eq. (\ref{eq:dqpmTc}) is obtained by requiring a constant energy density $\epsilon$ for the system along the critical line $T = T_c(\mu_q)$ where $\epsilon$ at $T_c(\mu_q = 0) \approx 0.158$ GeV is fixed by a lattice QCD calculation at $\mu_q = 0$. The coefficient in front of the $\mu_q^2$-dependent part can be compared to lQCD calculations at finite (but small) $\mu_B$ which gives \cite{Bonati:2014rfa} {\setlength\arraycolsep{0pt}
\begin{eqnarray}
 T_c (\mu_B) = T_c(\mu_B = 0) \cdot \biggl(1 - \kappa \left(\frac{\mu_B}{T_c (\mu_B = 0)}\right)^2 + \dots \biggr) \nonumber \\
\label{equ:Sec2.3}
\end{eqnarray}}
with $\kappa = 0.013(2)$. Rewriting (\ref{eq:dqpmTc}) in the form (\ref{equ:Sec2.3}) and using $\mu_B \approx 3 \mu_q$ we get $\kappa_{DQPM} \approx 0.0122$ which compares  well with the lQCD result \cite{Bonati:2014rfa}.
}

Thus, the DQPM effective coupling $\alpha_S(T,\mu_\mathrm{B})$ reads
\begin{equation}
\label{eq:as_DQPM}
\alpha_S(T,\mu_{\mathrm{B}}) \equiv
\begin{cases}
\dfrac{g^2(T,\mu_{\mathrm{B}}=0)}{4\pi},
& \mu_{\mathrm{B}} = 0, \\[6pt]
\dfrac{g^2\!\left(T_{\mathrm{scale}}(T,\mu_{\mathrm{B}})\right)}{4\pi},
& \mu_{\mathrm{B}} > 0,
\end{cases}
\end{equation}
with
\begin{equation}
T_{\mathrm{scale}} =
\frac{T^*}{T_C(\mu_{\mathrm{B}})/T_C(0)}. 
\end{equation}

\begin{figure}[h!]
  \includegraphics[width=\columnwidth]{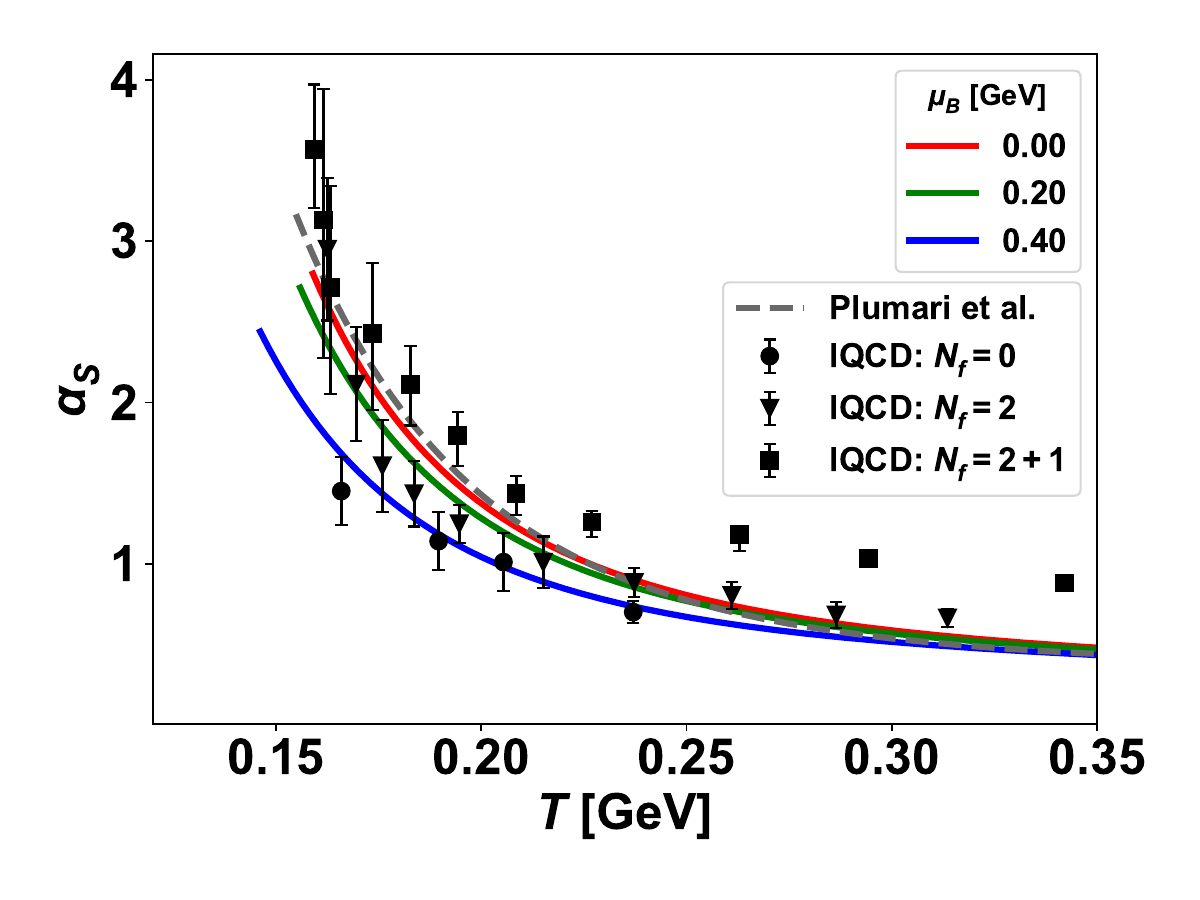}
  \caption{The strong coupling $\alpha_S$ as a function of  temperature $T$ at fixed $\mu_B =0.0, \ 0.2, \ 0.4$ GeV in comparison to the quasiparticle model from Ref.~\cite{Plumari:2011mk} and lQCD data.
The lattice results for quenched QCD, $N_f = 0$, (circles) are taken from Ref.~\cite{Kaczmarek:2004gv}, for $N_f = 2$ (inverted triangles) -- from Ref.~\cite{Kaczmarek:2005ui} and for $N_f =2 +1$ (squares) -- from Ref.~\cite{Kaczmarek:2007pb}.}
  \label{fig:alpha}
\end{figure}

\EB{ Figure~\ref{fig:alpha} shows the strong coupling $\alpha_s$ as a function of temperature $T$ for fixed baryon chemical potentials $\mu_B=0,\ 0.2,$ and $0.4$~GeV, together with the corresponding lattice-QCD results from Refs.~\cite{Kaczmarek:2004gv,Kaczmarek:2005ui,Kaczmarek:2007pb}. For comparison, we also display the coupling obtained within the quasiparticle model of Ref.~\cite{Plumari:2011mk}, which employs an explicit parametrization of $\alpha_s(T)$ similar to that used in earlier versions of the DQPM~\cite{Cassing:2007nb}. The two prescriptions---extracting $\alpha_s(T,\mu_B)$ from the lattice-QCD entropy density, as implemented in the present DQPM, and adopting an explicit temperature-dependent parametrization---lead to a qualitatively similar behavior. This agreement reflects the fact that both quasiparticle approaches are constrained to reproduce the lattice-QCD entropy density.
}

\subsection{Matrix elements in the DQPM}
\label{sec:DQPM_ME}
To evaluate the interaction rates of partons, one needs to compute the scattering amplitudes for all processes in which the corresponding parton species can participate.
\GI{The matrix elements for the $2\rightarrow2$ and forward $2\rightarrow3$ processes are evaluated from the corresponding leading-order Feynman diagrams, including the relevant s-, t-, and u-channel contributions and their interference terms. The inverse $3\rightarrow2$ matrix elements are related to the corresponding forward ones by microreversibility.}

In order to calculate the matrix elements corresponding to a scattering of DQPM partons, the scalar propagators have to be replaced by the propagators -- with full Lorentz structure -- for a massive (vector) gluon and massive (spin $1/2$) fermion with a finite width \cite{Berrehrah:2013mua}. 
In particular, the effective propagators and vertices entering the scattering amplitudes are given by

\begin{align}
  & \GP{\mu}{\nu}{p} =
  \left[-i\frac{g_{\mu\nu}-(p_\mu p_\nu) / m_g^2}
  {p^2-m_g^2+2 i \gamma_g \omega_p}\right]
   \text{(gluon propagator)},
  \nonumber\\
  &\QP{p} =
  \left[i \frac{\slashed{p}+m_q}
  {p^2-m_q^2 + 2i\gamma_q \omega_{p}}\right]
  \hspace{0.7cm} \text{(quark propagator)},
  \nonumber\\
  &\VX{\nu}{a}{ik} =
  \left(-ig\gamma^{\nu} T^a_{ik}\right)
  \hspace{2.2cm} \text{(quark-gluon vertex)},
  \nonumber\\
  & \GV{abc}{\mu\nu\tau}{p_1,p_2,p_3}
   =
  -g f^{abc} C^{\mu\nu\tau}(p_1,p_2,p_3)
  \nonumber\\
  &\hspace{5.1cm} \text{(three-gluon vertex)} .
\end{align}

Here $\mu,\nu,\tau$ denote Lorentz indices, $i,k=1,\dots,3$ are quark color indices, and $a,b,c=1,\dots,8$ are gluon color indices. 
In the DQPM, only transverse gluons are included in the calculation of thermodynamic quantities, in that the contribution of hard longitudinal gluons is found to be negligible \cite{Blaizot:2000fc,Peshier:2000hx}. 

\subsubsection*{$2\rightarrow 2$ scattering processes}

In this work we include all possible $2 \rightarrow 2$ partonic processes among light (anti)quarks, strange (anti)quarks, and gluons. \GI{When computing the microscopic interaction rate (Sec.~\ref{sec:TC}) of a $u$ quark, the processes considered are:
\begin{align}
    u+q &\rightarrow u+q\nonumber\\ 
    u+\bar q &\rightarrow u+\bar q\nonumber\\
    u+\bar u &\rightarrow q+\bar q\nonumber\\
    u+\bar u &\rightarrow g+ g\nonumber\\
    u+g &\rightarrow u+g
    \label{eq:elastic_u_processes}
\end{align} 
while for a gluon:
\begin{align}
    g+q &\rightarrow g+q\nonumber\\ 
    g+\bar q &\rightarrow g+\bar q\nonumber\\ 
    g+g &\rightarrow q+\bar q\nonumber\\
    g+g &\rightarrow g+g
    \label{eq:elastic_g_processes}
\end{align}}
with $q=u,d,s$ and $\bar q=\bar u,\bar d,\bar s$.

\begin{figure}[t!]
  \includegraphics[width=0.24 \columnwidth ]{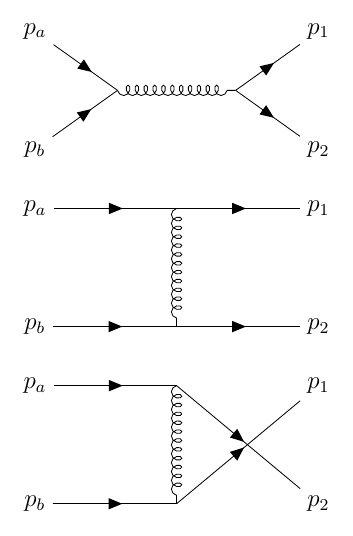}
  \includegraphics[width=0.48\columnwidth]{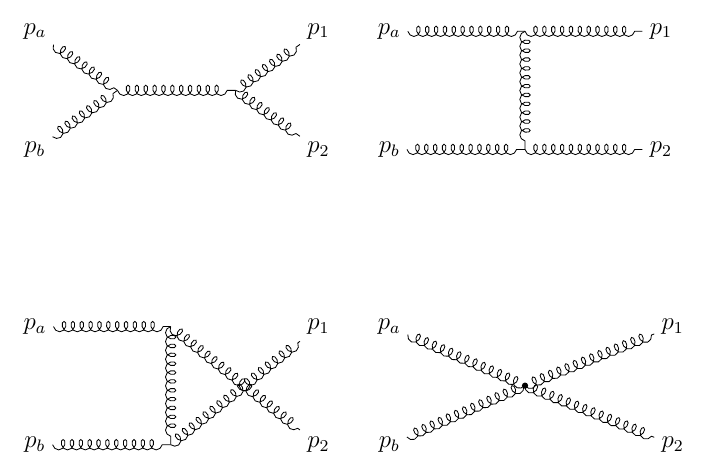} 
  \includegraphics[width=0.24 \columnwidth ]{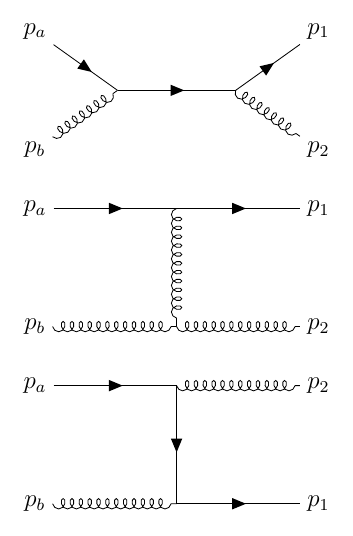}
  \caption{Leading-order Feynman diagrams for $q+q\rightarrow q+q$ (left column), $g+g\rightarrow g+g$ (middle columns), and $q+g\rightarrow q+g$ (right column) processes.}
  \label{fig:diag_22}
\end{figure}

Further details on the calculation of the DQPM invariant amplitudes for $2\rightarrow 2$ scattering channels -- shown in Fig.~\ref{fig:diag_22} --  
can be found in  Refs.~\cite{Moreau:2019vhw,Soloveva:2020hpr,Grishmanovskii:2022tpb}.

\begin{figure}[th!]
    \centering
    \includegraphics[width=0.96\columnwidth ]{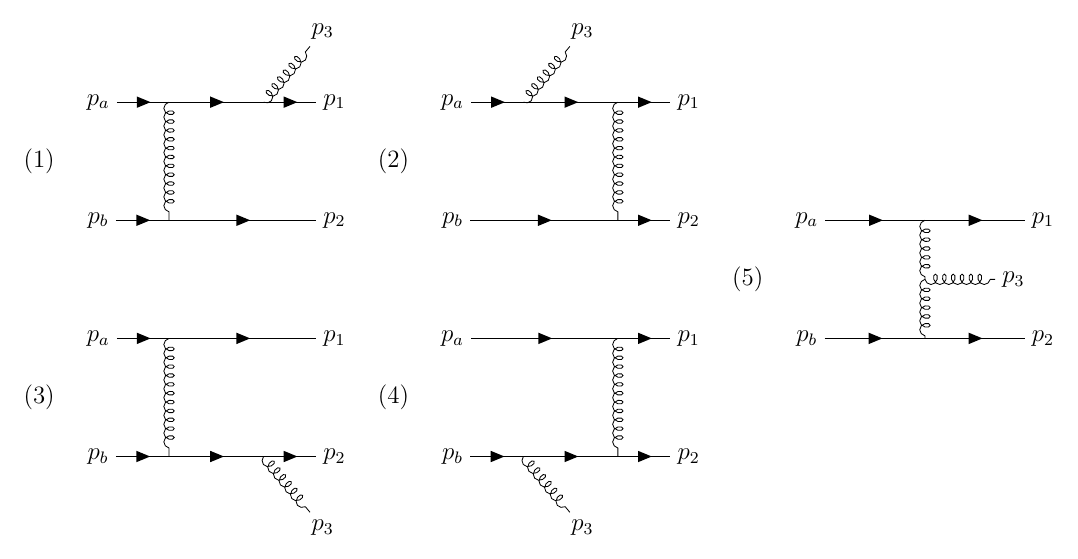}
    \caption{ Leading-order Feynman diagrams for $q+q \rightarrow q+q+ g$ processes in the $t$-channel. }
\label{fig:diags_23qq}
\end{figure}

\begin{figure}[th!]
    \centering
    \includegraphics[width=0.98\columnwidth]{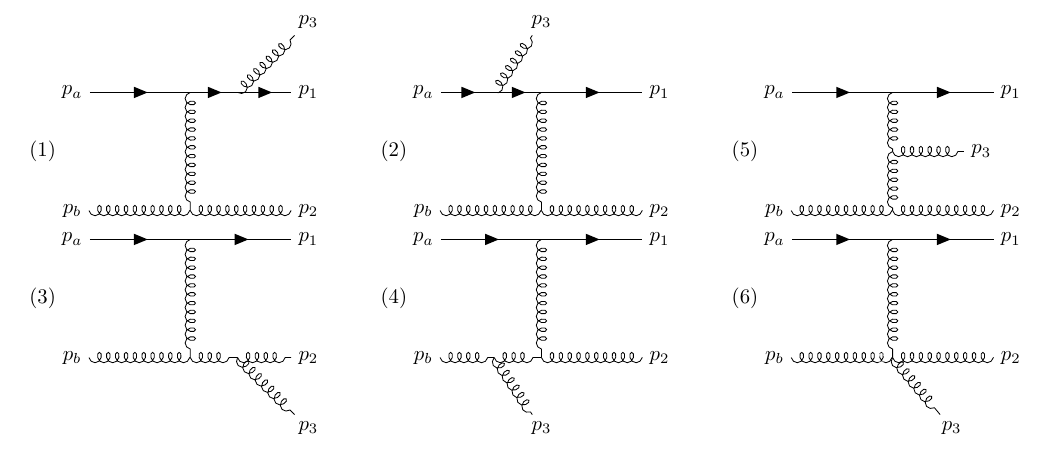}
    \caption{
      Leading-order Feynman diagrams for $q+g \rightarrow q+g+ g$ processes in the $t$-channel. 
    }
\label{fig:diags_23qg}
\end{figure}

\subsubsection*{$2 \leftrightarrow 3$ scattering processes}

For the $2\rightarrow3$ sector we consider gluon-radiation processes in which the incoming partons retain their identity and an additional final-state gluon is emitted.
\GI{The corresponding inverse $3\to2$ gluon-absorption matrix elements are related to the corresponding forward ones by microreversibility.}

The inelastic processes entering the microscopic interaction rate of a $u$ quark are:
\begin{align}
    u+q &\leftrightarrow u+q+g\nonumber\\ 
    u+\bar q &\leftrightarrow u+\bar q+g\nonumber\\ 
    u+g &\leftrightarrow u+g+g
    \label{eq:inelastic_u_processes}
\end{align}

similarly, for a gluon:

\begin{align}
    g+q &\leftrightarrow g+q+g\nonumber\\
    g+\bar q &\leftrightarrow g+\bar q+g\nonumber\\ 
    g+g &\leftrightarrow g+g+g.
    \label{eq:inelastic_g_processes}
\end{align}

In the following we give an overview of the DQPM invariant amplitudes calculations for $2\rightarrow3$ gluon-radiation processes, more details are given in Refs.~\cite{Grishmanovskii:2023gog,Grishmanovskii:2024gag}.
\EB{For clarity and compactness, Fig.~\ref{fig:diags_23qq} and \ref{fig:diags_23qg}  only display the Feynman diagrams corresponding to the $t$-channel, which gives the dominant contribution. The $u$- and $s$-channel contributions are evaluated analogously and are included in the numerical results.}

We start by considering gluon-radiation processes from $q+q$ scattering.
Using the notation above, we can write the invariant amplitudes $i \mathcal{M}_i = i \mathcal{M}_i (q^i q^k \rightarrow q^j q^l g^b)$ corresponding to each process in the following way:

\begin{widetext}
\begin{align}
    i \mathcal{M}_1 &=
    \ubar^l(p_2) \VX{\nu}{a}{lk} u^k(p_b) \GP{\mu}{\nu}{p_b-p_2}
    \ubar^j(p_1) \varepsilon_{\tau}^{*}(p_3) \VX{\tau}{b}{jm}
    \QP{p_1+p_3}
    \VX{\mu}{a}{mi} u^i(p_a),
    \nonumber\\[1pt]
    i \mathcal{M}_2 &=
    \ubar^j(p_1) \VX{\mu}{a}{ji} u^i(p_a) \GP{\mu}{\nu}{p_a-p_1}
    \ubar^l(p_2) \varepsilon_{\tau}^{*}(p_3) \VX{\tau}{b}{lm}
    \QP{p_2+p_3}
    \VX{\nu}{a}{mk} u^k(p_b),
    \nonumber\\[1pt]
    i \mathcal{M}_3 &=
    \ubar^l(p_2) \VX{\nu}{a}{lk} u^k(p_b) \GP{\mu}{\nu}{p_b-p_2}
    \ubar^j(p_1) \VX{\mu}{a}{jm}
    \QP{p_a-p_3}
    \varepsilon_{\tau}^{*}(p_3) \VX{\tau}{b}{mi} u^i(p_a),
    \nonumber\\[1pt]
    i \mathcal{M}_4 &=
    \ubar^j(p_1) \VX{\mu}{a}{ji} u^i(p_a) \GP{\mu}{\nu}{p_a-p_1}
    \ubar^l(p_2) \VX{\nu}{a}{lm} 
    \QP{p_b-p_3}
    \varepsilon_{\tau}^{*}(p_3) \VX{\tau}{b}{mk} u^k(p_b),
    \nonumber\\[1pt]
    i \mathcal{M}_5 &=
    \ubar^j(p_1) \VX{\mu}{a}{ji} u^i(p_a) \; \ubar^l(p_2) \VX{\lambda}{c}{lk} u^k(p_b)
    \GP{\mu}{\nu}{p_a-p_1}
    \GP{\lambda}{\sigma}{p_b-p_2} \GL{*}{\tau}{p_3}
    \nonumber \\ &\hspace{7cm}\times
    \GV{abc}{\sigma\tau\nu}{p_b-p_2,-p_3,p_2-p_b+p_3}.
    \label{eq:M}
\end{align}
\end{widetext}

Then the total invariant amplitude squared, averaged over initial states of spin and color, and summed over final states, for the $q+q$ scattering reads

\begin{align}
    & \Msq{qq' \rightarrow qq'g} = \frac{1}{N_c^2} \sum_{\text{color}}\frac{1}{(2s_q+1)(2s_{q'}+1)} \nonumber \\
    & \ \ \times \sum_{\text{spin}}
    |\mathcal{M}_1 + \mathcal{M}_2 + \mathcal{M}_3 + \mathcal{M}_4 + \mathcal{M}_5|^2,
    \label{eq:Mbar_qq}
\end{align}
where $q'$ denotes a quark with a possible different flavor than the $q$ quark.

Analogously one can compute the invariant $t$-channel amplitudes for gluon-radiation processes from $q+g$ scattering - presented in Fig.~\ref{fig:diags_23qg} - as:

\begin{widetext}    
\begin{align}
    i \mathcal{M}_1 &= 
    \ubar(p_1) 
    \VX{\mu}{d}{kn} 
    \QP{p_1+p_3, m_q} 
    \VX{\lambda}{f}{ni} 
    u(p_a)
    \GV{bef}{\sigma\tau\nu}{p_b-p_2,-p_b,p_2}
    \nonumber\\ & \hspace{7cm}\times
    \GP{\lambda}{\sigma}{p_2-p_b} \GL{*}{\mu}{p_3} \GL{*}{\nu}{p_2} \GL{}{\tau}{p_b},
    \nonumber \\[1pt]
    i \mathcal{M}_2 &= 
    \ubar(p_1) 
    \VX{\sigma}{f}{kn} 
    \QP{p_1+p_2-p_b, m_q} 
    \VX{\mu}{d}{ni} 
    u(p_a)
    \GV{bef}{\lambda\tau\nu}{p_b-p_2,-p_b,p_2}
    \nonumber\\ & \hspace{7cm}\times
    \GP{\lambda}{\sigma}{p_2-p_b} \GL{*}{\mu}{p_3} \GL{*}{\nu}{p_2} \GL{}{\tau}{p_b},
    \nonumber\\[1pt]
    i \mathcal{M}_3 &= 
    \ubar(p_1) 
    \VX{\lambda}{f}{ki} 
    u(p_a) 
    \GV{bfn}{\nu\mu\xi}{-p_2,-p_3,p_2+p_3}
    \GV{deh}{\sigma\tau\rho}{-p_b+p_2+p_3,p_b,-p_2-p_3}
    \nonumber\\ & \hspace{7cm}\times
    \GP{\rho}{\xi}{p_2-p_b} 
    \GP{\lambda}{\sigma}{p_2-p_b}
    \GL{*}{\mu}{p_3} \GL{*}{\nu}{p_2} \GL{}{\tau}{p_b},
    \nonumber\\[1pt]
    i \mathcal{M}_4 &= 
    \ubar(p_1) 
    \VX{\lambda}{f}{ki} 
    u(p_a) 
    \GV{efh}{\tau\mu\rho}{-p_b,p_3,p_b-p_3}
    \GV{bdh}{\nu\sigma\xi}{p_2,p_b-p_2-p_3,-p_b+p_3}
    \nonumber\\ &\hspace{7cm}\times
    \GP{\rho}{\xi}{p_2-p_b} 
    \GP{\lambda}{\sigma}{p_2-p_b} \GL{*}{\mu}{p_3} \GL{*}{\nu}{p_2} \GL{}{\tau}{p_b},
    \nonumber\\[1pt]
    i \mathcal{M}_5 &= 
    \ubar(p_1) 
    \VX{\lambda}{f}{ki} 
    u(p_a) 
    \GV{dfh}{\sigma\mu\xi}{-p_b+p_2+p_3,-p_3,-p_2+p_b}
    \GV{beh}{\nu\tau\rho}{-p_2,p_b,-p_b+p_2}
    \nonumber\\ &\hspace{7cm}\times
    \GP{\rho}{\xi}{p_2-p_b} 
    \GP{\lambda}{\sigma}{p_2-p_b} \GL{*}{\mu}{p_3} \GL{*}{\nu}{p_2} \GL{}{\tau}{p_b},
    \nonumber\\[1pt]
    i \mathcal{M}_6 &= 
    \ubar(p_1) 
    \VX{\lambda}{f}{ki} 
    u(p_a) \,
    \Gamma^{fbde}_{\mu\nu\sigma\tau} \,
    \GP{\lambda}{\sigma}{p_2-p_b} \GL{*}{\mu}{p_3} \GL{*}{\nu}{p_2} \GL{}{\tau}{p_b}. 
    \label{eq:Mglue}
\end{align}
\end{widetext}

In this case the averaged squared amplitude reads
\begin{multline}
    \Msq{qg \rightarrow qgg} = \frac{1}{N_c(N_c^2-1)} \sum_{\text{color}}\frac{1}{(2s_q+1)d_g}
    \\ \times
    \sum_{\text{spin}}
    |\mathcal{M}_1 + \mathcal{M}_2 + \mathcal{M}_3 + \mathcal{M}_4 + \mathcal{M}_5 + \mathcal{M}_6|^2,
    \label{eq:Mbar_qg}
\end{multline}
where $d_g$ is the degeneracy factor of the gluon. 
In addition, the $gg \rightarrow ggg$ process, which includes in total 25 diagrams, has been evaluated employing the scaling relation $\sigma_{gg \rightarrow ggg} = \frac{9}{4}\sigma_{qg \rightarrow qgg}$ for simplicity \cite{Grishmanovskii:2024qtm}.

\section{Transport coefficients} 
\label{sec:TC}
In this section, we focus on the near equilibrium transport coefficients of the QGP within the DQPM such as the shear viscosity $\eta$, bulk viscosity $\zeta$, electric conductivity $\sigma_Q$, and baryon diffusion coefficient $\kappa_B$.

It has been shown that, within the quasiparticle approximation, transport coefficients obtained from the relaxation-time approximation (RTA) of kinetic theory \cite{Hosoya:1983xm,Chakraborty:2010fr,Albright:2015fpa,Gavin:1985ph} are numerically very close to those extracted from one-loop Kubo-type calculations \cite{Kubo:1957mj,Aarts:2002,Fernandez-Fraile:2005bew}. Accordingly, all transport coefficients discussed below are evaluated within the RTA framework.

\subsection{Relaxation times}

The first step in the RTA calculation is the determination of the relaxation times, which in general depend on momentum, temperature, and baryon chemical potential.
The relaxation times for quarks and gluons are given by:
\begin{align}
\tau_i(\mathbf{p},T,\mu_B) = \frac{1}{\Gamma_i(\mathbf{p},T,\mu_B)} 
\end{align}
where $\Gamma_i(\mathbf{p},T,\mu_B)$ is the parton interaction rate, whose relation to the microscopic DQPM scattering amplitudes is made explicit in the following subsection.

\GI{
We emphasize that in the present work $\tau_i$ represents a microscopic single-particle relaxation time within kinetic theory. It should not be confused with second-order hydrodynamic relaxation times $\tau_\pi$ and $\tau_\Pi$, which govern the transient evolution of the shear-stress tensor and bulk viscous pressure \cite{Israel:1979wp,Denicol:2012cn,Huang:2011dc}, respectively, and are not evaluated here.

Within the scalar RTA, the same microscopic relaxation time is employed in the tensor, scalar, and vector sectors, providing a unified connection between the underlying collision dynamics and the different transport coefficients. These coefficients remain distinct, as they involve different momentum moments and thermodynamic source terms, although their dependence on the collision model is correlated through the common function $\tau_i(\mathbf{p},T,\mu_B)$.

A more differential description may associate distinct relaxation scales with the different eigenmodes of the linearized collision operator. Such extensions can be developed through Chapman--Enskog or moment-based approaches \cite{Fotakis:2019nbq,Parisi:2026ufk}. More generally, mode-resolved approaches replace a single relaxation rate with distinct rates for the relevant collision-operator eigenmodes \cite{Soloveva:2026MRTA}. Implementation of such a mode-resolved treatment in the DQPM framework is left for future work.
}

We then compare the resulting transport coefficients in order to quantify the impact of inelastic $2\leftrightarrow3$ channels within this RTA implementation.

\subsubsection{$2\rightarrow2$ interaction rates}

We briefly recall that for on-shell partons (with energies taken to be $E^2 = \mathbf{p}^2 + m^2$ where $m$ is the pole mass) the $2\rightarrow2$ interaction rate (i.e. collisional width) can be computed as follows:
\begin{align}
&\Gamma^{ 2 \rightarrow 2}_i (\mathbf{p}_i, T,\mu_B) = \nonumber \\
& \ \ = \frac{1}{2E_i} \sum_{j=q,\bar{q},g} \int \frac{d^3{\bf p}_j}{(2\pi)^3 2E_j}\ d_j\ f_j(E_j,T,\mu_q)  \nonumber \\
& \ \ \times  \int \frac{d^3{\bf p}_1}{(2\pi)^3 2E_1}  \int \frac{d^3{\bf p}_2}{(2\pi)^3 2E_2} (1\pm f_1) (1\pm f_2) \nonumber \\[2pt]
& \ \ \times |\bar{\mathcal{M}}_{ 2 \rightarrow 2}|^2 \ (2\pi)^4 \delta^{(4)}\left(p_i + p_j -p_1 -p_2 \right) .
\label{Gamma_22}
\end{align}

Here and in the following, $d_j$ stand for the degeneracy factors for spin and color ($d_q = 2 \times N_c$ for quarks and $d_g =2 \times (N_c^2-1)$ for gluons), and we adopt the shorthand notation $f_j = f_j(E_j,T,\mu_q)$ for the equilibrium distribution functions. Moreover, the Pauli-blocking $(1-f_k)$ and Bose-enhancement $(1+f_k)$ factors account for the available density of final states, and the notation $\sum_{j=q,\bar{q},g}$ includes the contribution from all partons, i.e. the gluons and the (anti)quarks of three different flavors ($u,d,s$).

One can compute the average interaction rate of parton $i$ over its momentum distribution as
\begin{align}
\Gamma^{ 2 \rightarrow 2}_i(T,\mu_B) = \frac{d_i}{n_i(T,\mu_q)} \int \frac{d^3{\bf p}_i}{(2\pi)^3}\ f_i(E_i,T,\mu_q)  \nonumber \\ 
 \times \ \Gamma^{ 2 \rightarrow 2}_i(\mathbf{p}_i,T,\mu_B) \ ,
\label{Gamma_22_av}
\end{align}
with the on-shell density of parton $i$ at  $T$ and $\mu_q$
\begin{equation}
n_i(T,\mu_q) = d_i \int \frac{d^3{\bf p}_i}{(2\pi)^3}\  f_i(E_i,T,\mu_q) .
\label{eq:n_on}
\end{equation}

\subsubsection{$2\leftrightarrow3$ interaction rates}
\GI{We evaluate separately the forward $2\to3$ and inverse $3\to2$ interaction rates using the time-reversed matrix elements and the appropriate thermal phase-space and statistical factors; detailed balance is used to verify the consistency of their implementation.\\}
\begin{figure}[t!]
    \centering
    \includegraphics[width=0.98\columnwidth]{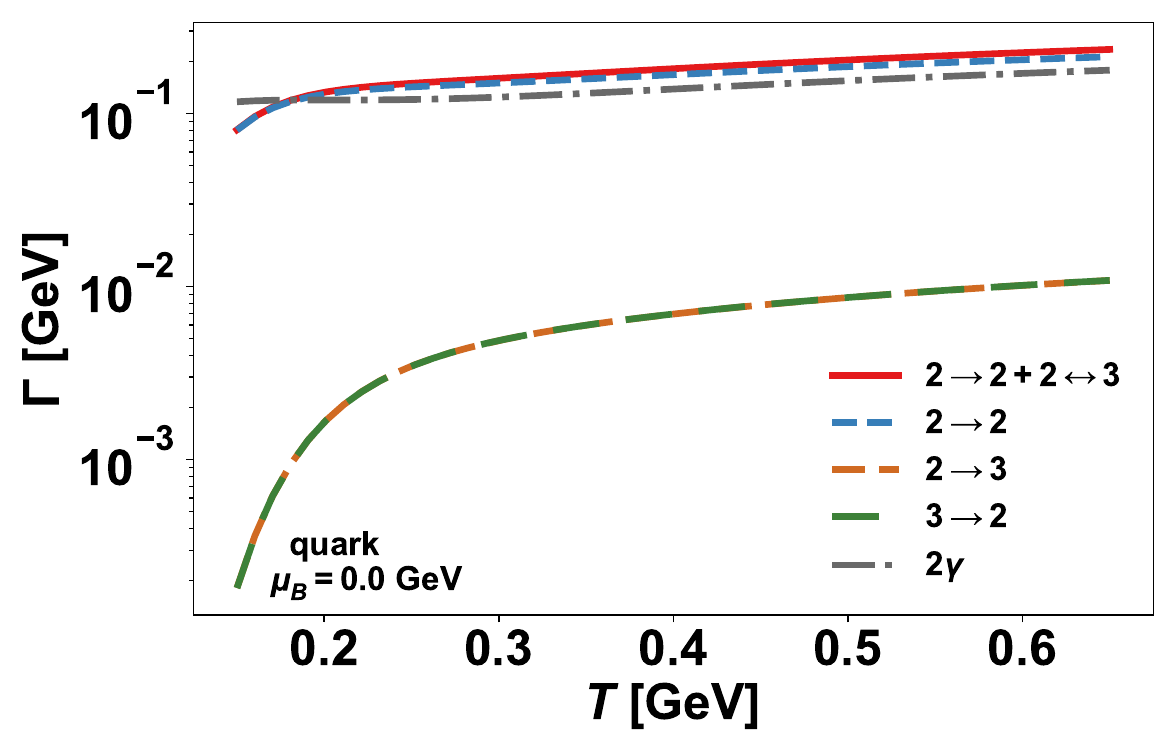}
    \includegraphics[width=0.98\columnwidth]{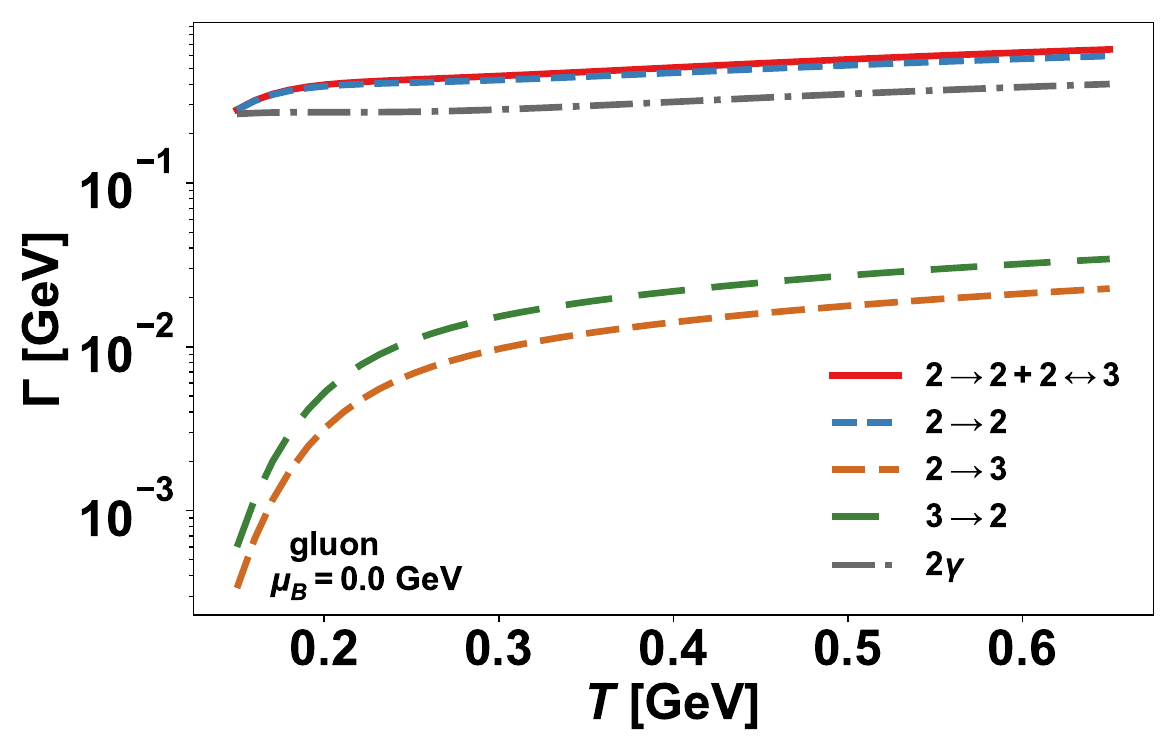}
    \caption{ Momentum averaged on-shell interaction rates $\Gamma(T,\mu_B)$ for a light quark (upper plot) and a gluon (lower plot) as a function of temperature $T$, at fixed $\mu_B = 0$, \GI{for $2\rightarrow2$ (dashed blue lines), $2\rightarrow3$ (dashed orange lines) and $3\rightarrow2$ (dashed green lines) processes; the sum of the three contributions is also displayed (solid red line).} Twice the DQPM effective widths $2\gamma$ (dotted gray lines) are shown for comparison. }
    \label{fig:IR_q_g}
\end{figure}
The $2\rightarrow3$ interaction rate for on-shell partons is evaluated by
\begin{align}
  \label{eq:Gamma_23}
  &\Gamma_i^{ 2\rightarrow 3}({\bf p}_i,T,\mu_B) = \nonumber\\
  & \ \ = \frac{1}{2E_i}\sum_{j=q,\bar{q},g}
   \int\frac{d^3{\bf p}_j}{\dpi^3 2E_j} d_j f_j \int\frac{d^3{\bf p}_1}{\dpi^3 2E_1}
  \nonumber\\
  & \ \ \times \int\frac{d^3{\bf p}_2}{\dpi^3 2E_2} \int\frac{d^3{\bf p}_g}{\dpi^3 2E_g} 
  (1 \pm f_1)(1 \pm f_2)(1 \pm f_g)
  \nonumber\\[2pt]
  & \ \ \times \ |\bar{\mathcal{M}}_{ 2 \rightarrow 3}|^2 \ (2\pi)^4 \delta^{(4)}(p_i + p_j - p_1 - p_2 - p_g),
\end{align}
\GI{
and the inverse $3\to2$ rate is analogously given by

\begin{align}
  \label{eq:Gamma_32}
  &\Gamma_i^{ 3\rightarrow 2}({\bf p}_i,T,\mu_B) = \nonumber\\
  & \ \ = \frac{1}{2E_i}\sum_{j=q,\bar{q},g}
   \int\frac{d^3{\bf p}_j}{\dpi^3 2E_j} d_j f_j \int\frac{d^3{\bf p}_g}{\dpi^3 2E_g} d_g f_g \nonumber\\ 
  & \ \ \times \int\frac{d^3{\bf p}_1}{\dpi^3 2E_1} \int\frac{d^3{\bf p}_2}{\dpi^3 2E_2}  
  (1 \pm f_1)(1 \pm f_2)
  \nonumber\\[2pt]
  & \ \ \times \ |\bar{\mathcal{M}}_{ 3 \rightarrow 2}|^2 \ (2\pi)^4 \delta^{(4)}(p_i + p_j + p_g - p_1 - p_2).
\end{align}
}
Additionally, the momentum averaged inelastic rates are computed as
\GI{
\begin{align}
  \Gamma_i^{a}(T,\mu_B) =& \frac{d_i}{n_i(T,\mu_q)} \int \frac{d^3{\bf p}_i}{\dpi^3} f_i(E_i,T,\mu_q) \Gamma_i^{a)}({\bf p}_i,T,\mu_B) \ ,
  \label{eq:GTmu}
\end{align}
where $a= 2\rightarrow 3 \ , 3 \rightarrow 2$ } 
and $n_i(T,\mu_q)$ is the on-shell density of parton $i$ at $T$ and $\mu_q$ as expressed in Eq. (\ref{eq:n_on}).\\


\begin{figure*}[th!]
    \centering
    \includegraphics[width=0.45\linewidth]{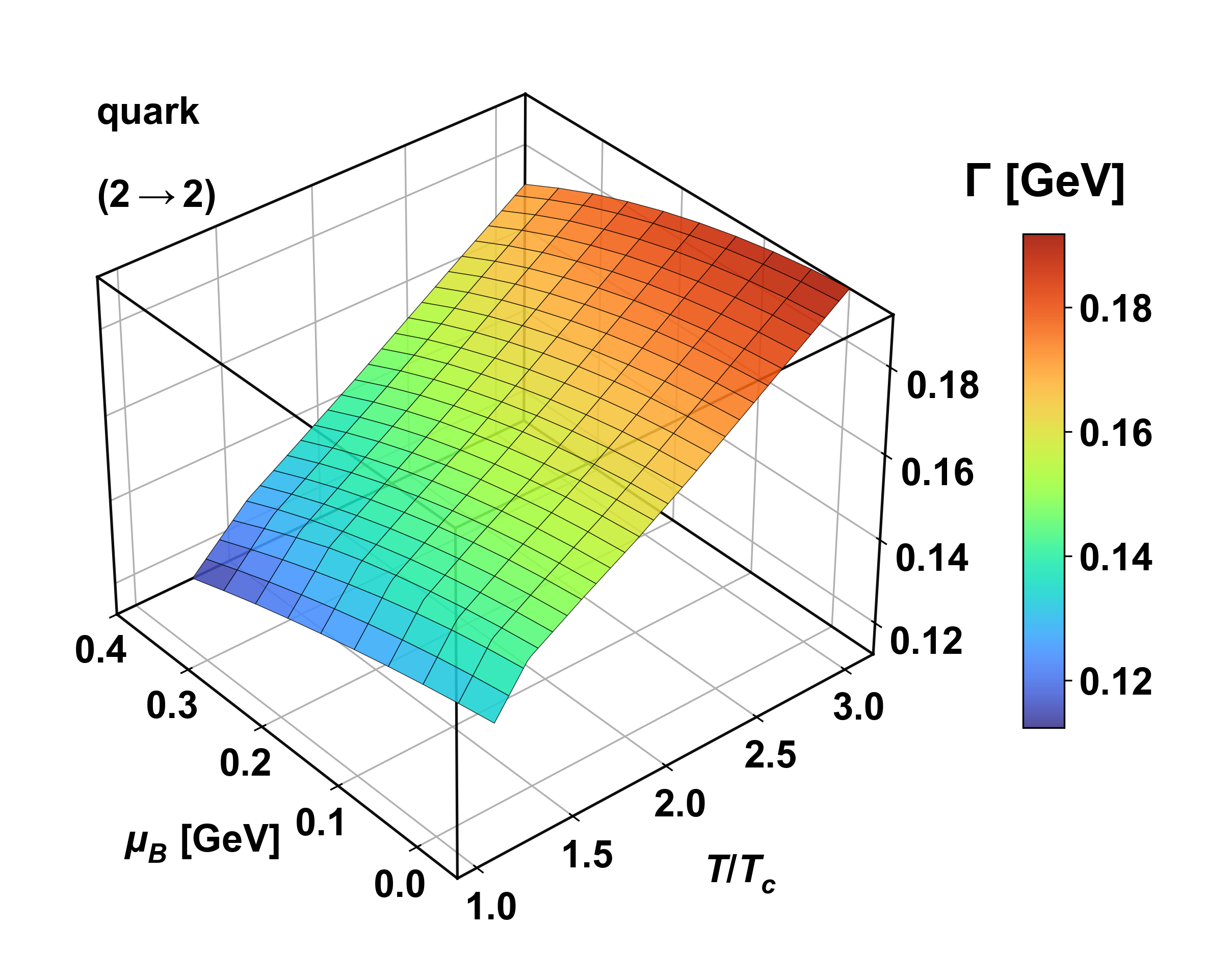}
    \includegraphics[width=0.45\linewidth]{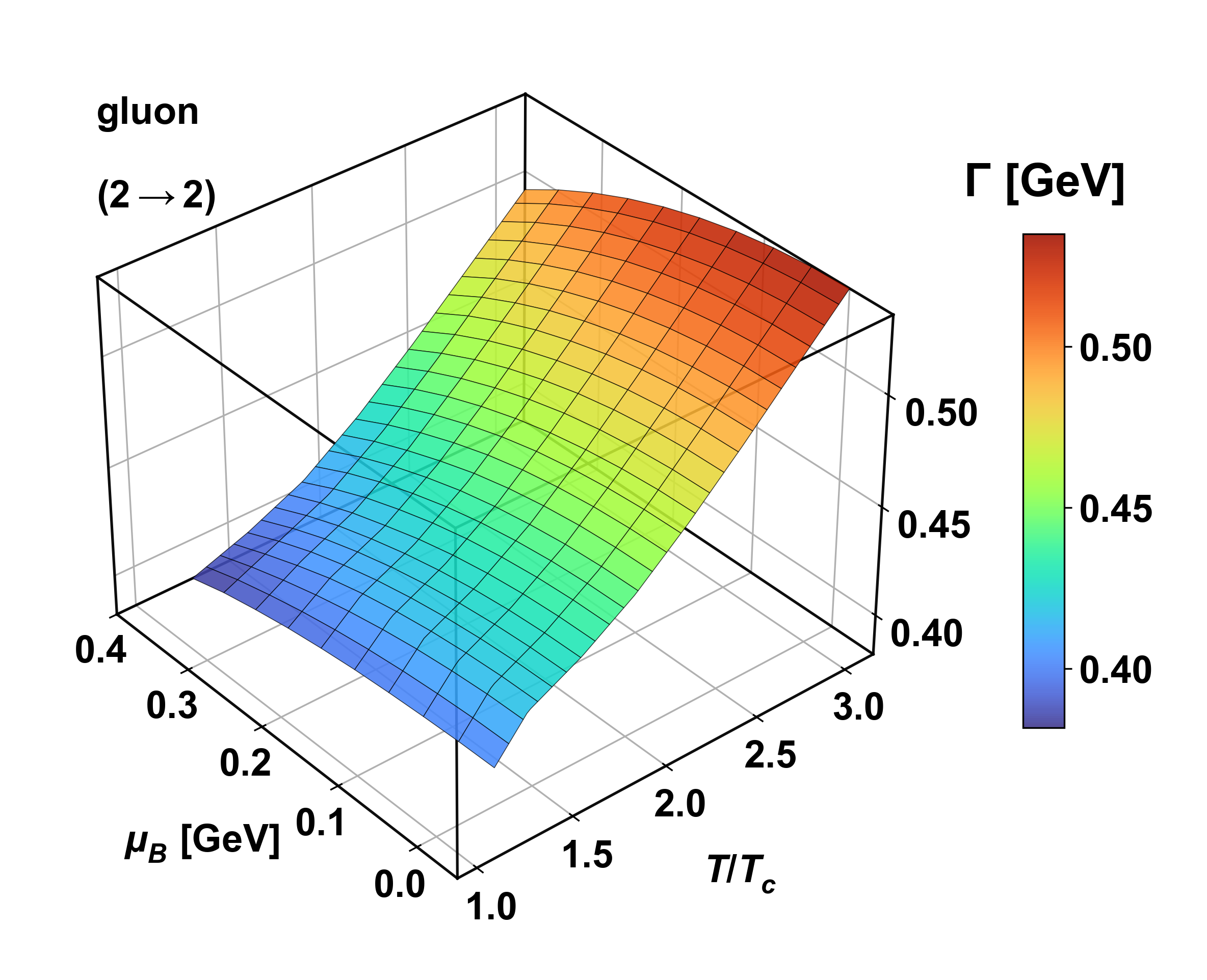}
    \includegraphics[width=0.45\linewidth]{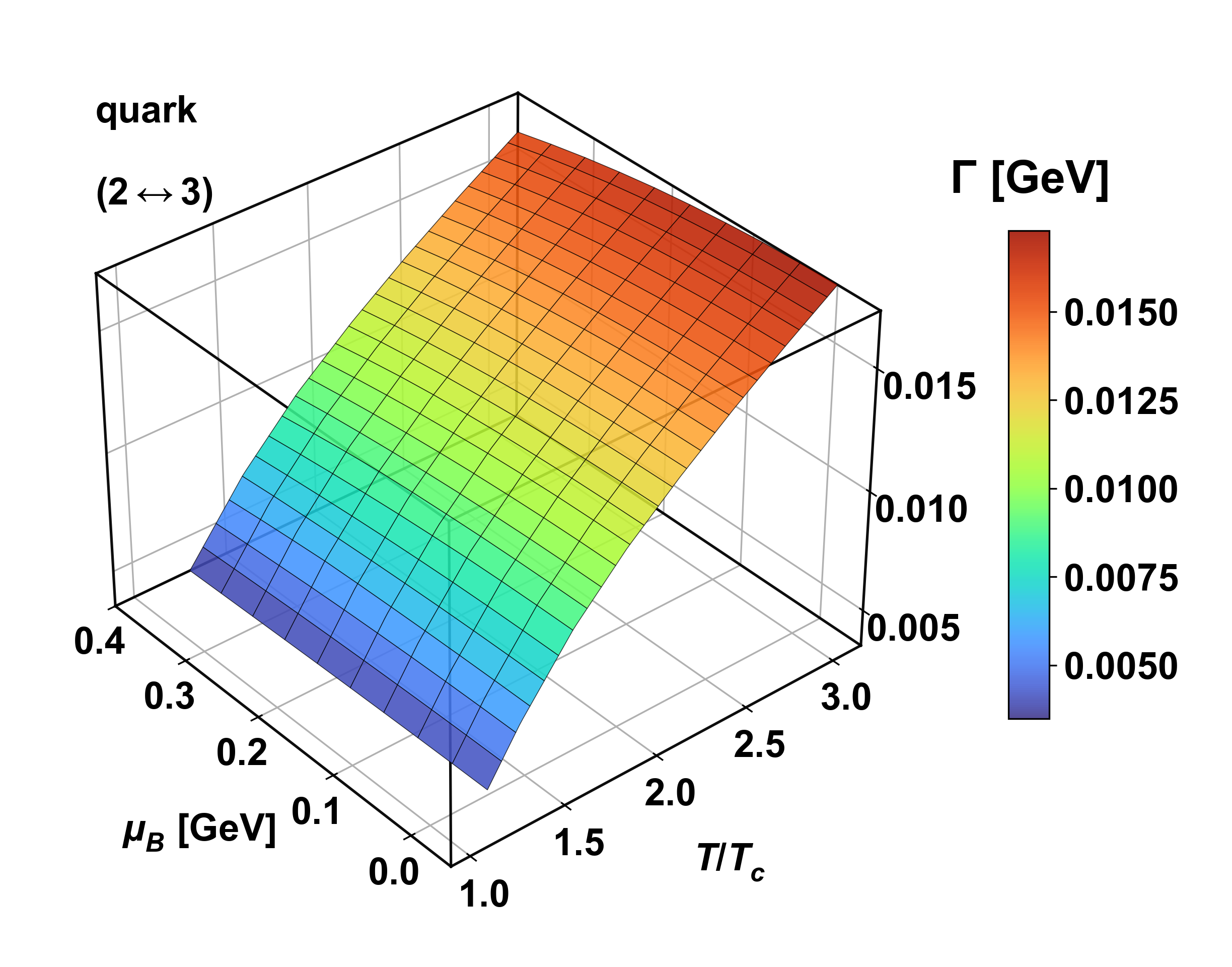}  
    \includegraphics[width=0.45\linewidth]{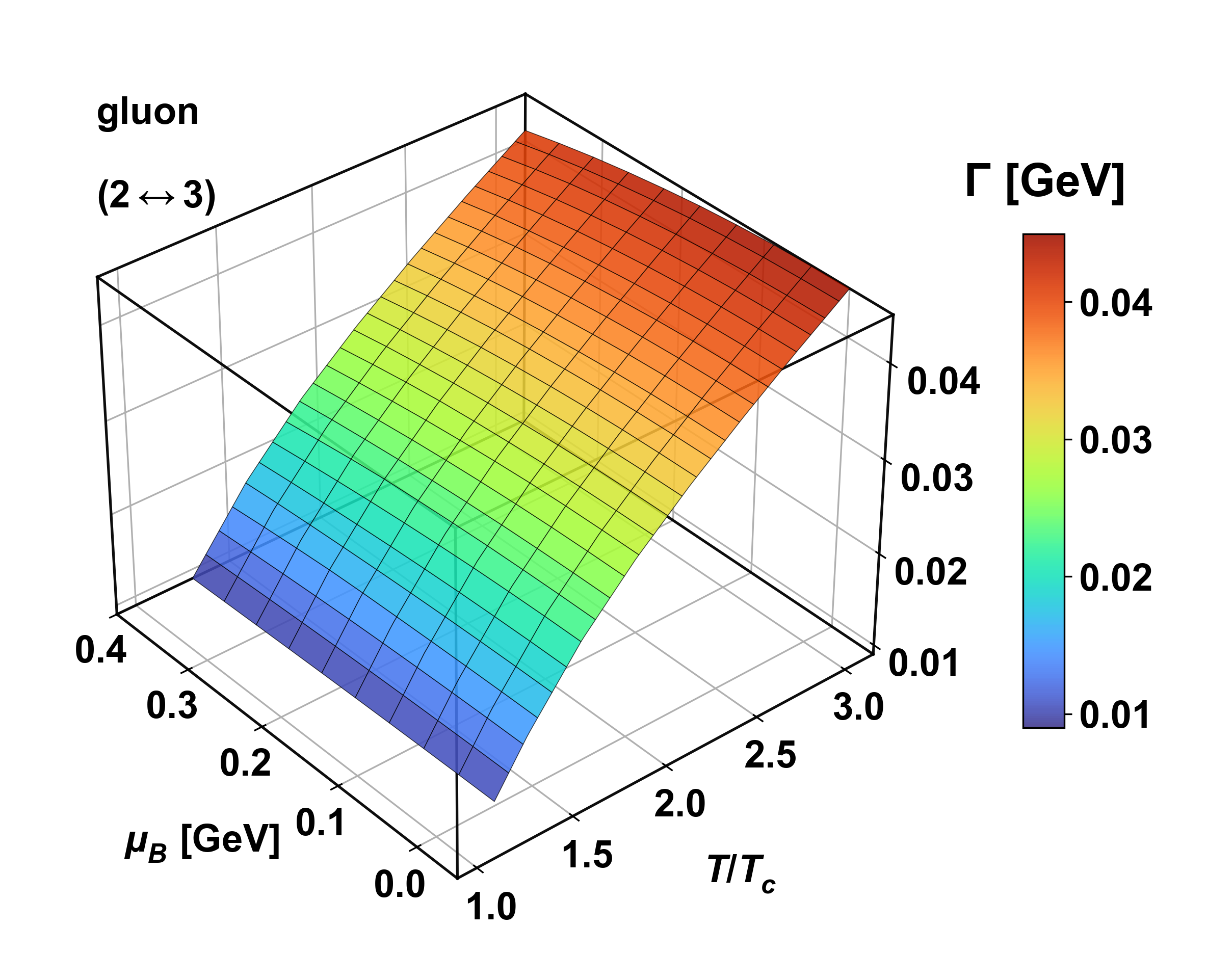}  
 \caption{Momentum averaged on-shell interaction rates $\Gamma(T,\mu_B)$ for a light quark (left column) and a gluon (right column) as a function of scaled temperature $T/T_C$ and baryon chemical potential $\mu_B$, for $2\rightarrow2$ (upper plots) and \GI{$2\leftrightarrow3$} (lower plots) processes. } 
  \label{fig:IR_3d}
\end{figure*}

Figure~\ref{fig:IR_q_g} shows the on-shell interaction rates for $2\rightarrow2$, \GI{ $2\rightarrow3$ and $3\rightarrow2$ reactions}, computed at fixed baryon chemical potential $\mu_B=0$, as a function of temperature $T$ and averaged over momentum, respectively for a light quark and for a gluon. \GI{Both the forward gluon-emission and inverse gluon-absorption interaction rates remain below the corresponding 2→2 rates over the temperature range considered.} This indicates that such inelastic channels provide a correction to the thermal relaxation times, and hence to the transport coefficients \GI{evaluated in RTA}, rather than a dominant contribution.

In addition, the $(T,\mu_B)$ dependence of momentum averaged interaction rates for partons is shown in Fig.~\ref{fig:IR_3d}. For both quarks and gluons, the $2\rightarrow2$ rates increase with temperature and show only a mild dependence on the baryon chemical potential in the range considered. The same qualitative behavior is observed for the inelastic $2\leftrightarrow3$ channels. Their rates increase with temperature and vary smoothly with $\mu_B$, but remain well below the corresponding $2\rightarrow2$ rates over the $(T,\mu_B)$ plane. This confirms that gluon-radiation and absorption processes provide a subleading correction to the total relaxation rate rather than a dominant contribution.

We emphasize that the averaged rates shown in Fig.~\ref{fig:IR_q_g} and Fig.~\ref{fig:IR_3d} are used only to illustrate the relative magnitude of the $2\rightarrow2$ and $2\leftrightarrow3$ contributions. In the calculation of the transport coefficients, the full momentum-dependent rates $\Gamma_i(\mathbf{p},T,\mu_B)$ are employed.

\EB{
We note that the DQPM is an effective quasiparticle approach constrained by lattice-QCD thermodynamics, with dressed propagators and vertices effectively encoding higher-order medium corrections. 
}
\GI{Although the inelastic $2\leftrightarrow3$ contributions are shown to be subleading relative to $2\rightarrow2$ scattering in the thermal regime considered here, more complex inelastic processes may become relevant at higher temperatures or for high-momentum probes, such as jets.}

\subsection{Shear and bulk viscosity}

Shear and bulk viscosities encode the dissipative response of the medium to gradients of the hydrodynamic flow. While shear viscosity $\eta\,(T,\mu_B)$ governs momentum diffusion and the damping of anisotropic flow, $\zeta\,(T,\mu_B)$ is sensitive to the breaking of conformal symmetry and is expected to be enhanced near the QCD crossover.

The shear and bulk viscosities for quasiparticles with medium-dependent masses $m_i(T,\mu_q)$ can be calculated by solving the linearized Boltzmann equation in the RTA \cite{Chakraborty:2010fr}, respectively, as
\begin{align}
\eta(T,\mu_B)  = \frac{1}{15T} \sum_{i=q,\bar{q},g} \int \frac{d^3{\bf p}}{(2\pi)^3} \frac{\mathbf{p}^4}{E_i^2}   \tau_i(\mathbf{p},T,\mu_B) \nonumber \\ 
\times  d_i f_i 
  \left(1\pm f_i\right) , 
\label{eq:eta_RTA}
\end{align}

and 

\begin{align}
  \zeta(T,\mu_B)= \frac{1}{9T} \sum_{i=q,\bar{q},g}\int \frac{d^3{\bf p}}{(2\pi)^3}\,    
  \frac{1}{E_i^2} \tau_i(\mathbf{p},T,\mu_B)  \nonumber \\[1pt]
 \times \left[\mathbf{p}^2-3c_s^2\left(E_i^2-T^2\frac{dm_i^2}{dT^2}\right)\right]^2 d_i f_i 
  \left(1\pm f_i\right),
  \label{eq:zeta_RTA}   
\end{align}
where $c_s^2$ is the speed of sound squared and $\tau_i$ is the relaxation time.\\

\begin{figure}[th!]
    \centering
    \includegraphics[width=\columnwidth]{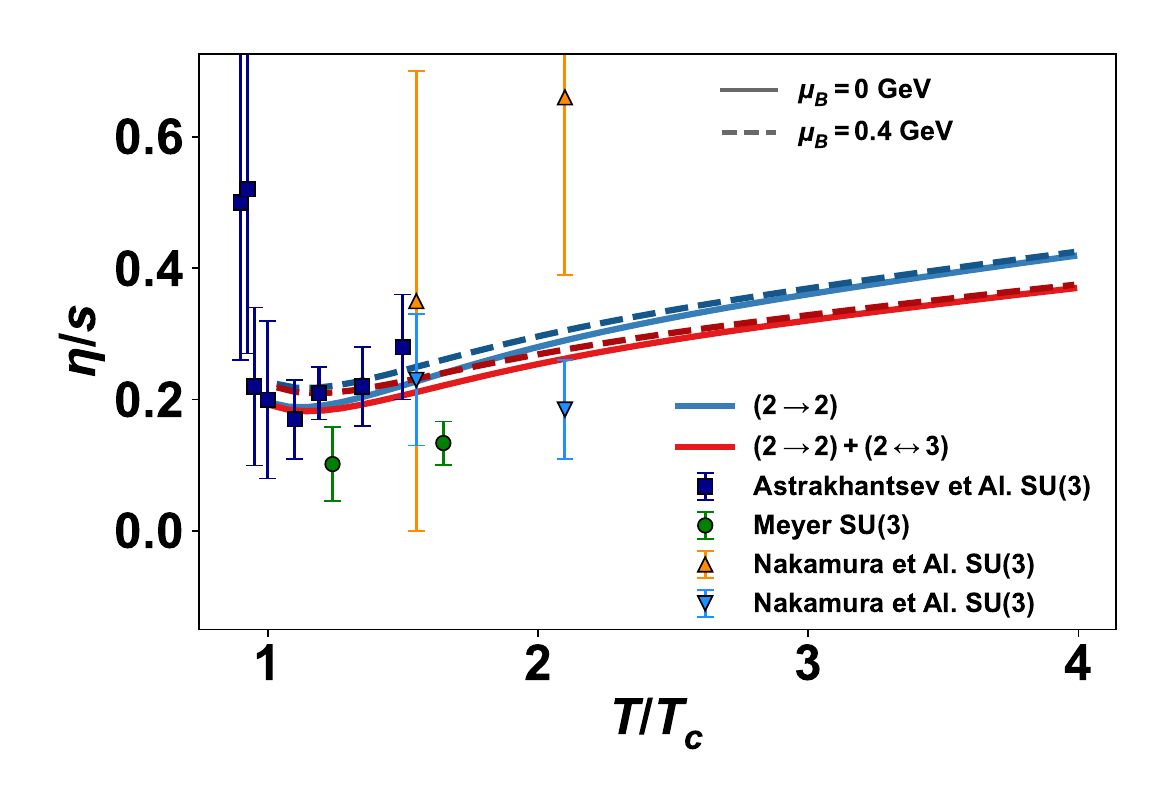}\\
    \caption{Shear viscosity to entropy density ratio $\eta/s$ as a function of scaled temperature $T/T_C$, at fixed values of baryon chemical potential \GI{$\mu_B=0$ (solid lines), $\mu_B=0.4$ GeV (dashed lines)}. Blue lines show $\eta/s$ for $2\rightarrow 2$ processes only, while red lines correspond to $\eta/s$ for both $2\rightarrow 2$ and $2\leftrightarrow3$ processes. Lattice results at $\mu_B=0$ for pure SU(3) gauge theory from Ref.~\cite{Astrakhantsev:2017nrs} (blue squares), Ref.~\cite{Meyer:2007ic} (green circles), Ref.~\cite{Nakamura:2004sy} (orange triangles), Ref.~\cite{Nakamura:2007nx} (cyan upside triangles) are shown.}
    \label{fig:TC_eta}
\end{figure}

The resulting shear viscosity to entropy density ratio $\eta/s$ is shown in Fig.~\ref{fig:TC_eta} as a function of the scaled temperature $T/T_C$ for $\mu_B=0$ and $\mu_B=0.4$ GeV. In both cases, the inclusion of $2\leftrightarrow3$ inelastic processes leads to a small reduction of $\eta/s$ with respect to the case of $2\rightarrow2$ scatterings alone. This is expected, since the additional inelastic channels increase the total interaction rate and therefore reduce the relaxation time entering Eq. (\ref{eq:eta_RTA}).

The moderate size of this effect is consistent with radiation/absorption processes becoming more relevant at large momenta, whereas thermal transport coefficients are dominated by momenta of order $p\sim\,T$. As a consequence, the $2\leftrightarrow3$ contribution acts as a correction to the $2\rightarrow2$ result rather than producing a qualitative modification of the temperature dependence of  $\eta/s$.

\begin{figure}[th!]
    \centering
    \includegraphics[width=\columnwidth]{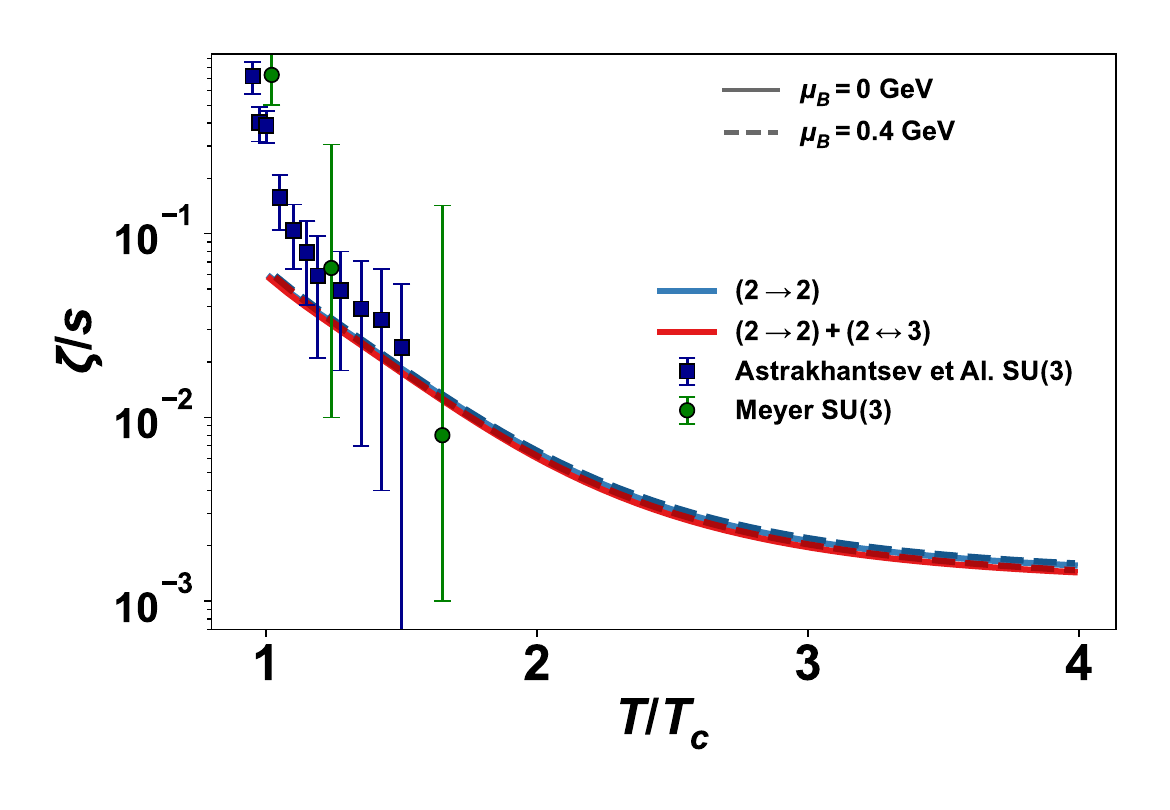}
    \caption{Bulk viscosity to entropy density ratio $\zeta/s$ as a function of scaled temperature $T/T_C$, at fixed values of baryon chemical potential \GI{$\mu_B=0$ (solid lines), $\mu_B=0.4$ GeV (dashed lines)}. Blue lines show $\zeta/s$ for $2\rightarrow 2$ processes only, while red lines correspond to $\zeta/s$ for both $2\rightarrow 2$ and $2\leftrightarrow3$ processes. Lattice results at $\mu_B=0$ for pure SU(3) gauge theory from Ref.~\cite{Astrakhantsev:2018oue} (blue squares), Ref.~\cite{Meyer:2007dy} (green circles) are shown.}
    \label{fig:TC_zeta}
\end{figure}

Similarly, the bulk viscosity to entropy density ratio $\zeta/s$ is shown in Fig.~\ref{fig:TC_zeta}. 
\GI{Within the RTA, the inclusion of inelastic processes has a very limited impact on $\zeta/s$}, since the integral in Eq. (\ref{eq:zeta_RTA}) is strongly weighted by the non-conformal factor {\small $\left[\mathbf{p}^2-3c_s^2\left(E_i^2-T^2\frac{dm_i^2}{dT^2}\right)\right]^2$}. This term determines the dominant momentum region contributing to $\zeta$, making the result less sensitive to the moderate modification of the relaxation time induced by inelastic processes. Consequently, the inclusion of $2\leftrightarrow3$ scatterings, entering only through the relaxation time, leaves the temperature dependence of $\zeta/s$ essentially unchanged and produces only a very small quantitative correction.\\

Moreover, at $\mu_B=0$, both bulk and shear viscosity results are compatible with the available lattice estimates for pure SU(3) gauge theory within uncertainties. Increasing the baryon chemical potential does not qualitatively change the temperature dependence of either quantity, while producing only small quantitative variations.


\subsection{Electric conductivity and baryon diffusion}

At large net-baryon density, diffusion of conserved charges can significantly affect the evolution of the medium. The corresponding transport coefficients are the diffusion coefficient $\kappa_a$, or equivalently the conductivity $\sigma_a=\kappa_a/T$, associated with a conserved charge $a=Q,B,S$, denoting electric charge, baryon number, and strangeness, respectively.

The electric conductivity $\sigma_Q(T,\mu_B)$ describes the response to a stationary external electric field and influences the time evolution of electromagnetic fields in heavy-ion collisions, with possible implications for the chiral magnetic effect. It also affects soft-photon emission rates \cite{Yin:2013kya} and spectra \cite{Turbide:2003si,Akamatsu:2011mw,Linnyk:2013hta}.

In the RTA, it is evaluated as \cite{Albright:2015fpa,Berrehrah:2014kba}:
\begin{align}
    \sigma_Q^{\mathrm{RTA}}(T,\mu_B) = \frac{e^2}{3T} \sum_{i=q,\bar q}
    q_i^2 \int \frac{d^3{\bf p}}{(2\pi)^3}&\frac{\mathbf{p}^2}{E_i^2}\tau_i(\mathbf{p},T,\mu_B) \nonumber\\ 
    &\times d_i f_i \left(1-f_i\right),
    \label{eq:sigma_Q_RTA}
\end{align}

where $q_i$ denotes the dimensionless electric charge of the parton species, namely
$q_u=2/3$ and $q_d=q_s=-1/3$, while antiquarks carry the opposite electric charge. The factor $e^2=4\pi\alpha_{\rm em}$ gives the electromagnetic coupling. Gluons do not contribute directly to Eq.~(\ref{eq:sigma_Q_RTA}), since they carry no electric charge; nevertheless, they affect $\sigma_Q$ indirectly through their contribution to the interaction rates entering the relaxation times of quarks and antiquarks.

The baryon diffusion coefficient $\kappa_B(T,\mu_B)$ describes the response of the net-baryon current to gradients of the baryon chemical potential. In the Landau frame, the diffusive current is defined orthogonally to the energy flow, leading to the projected baryon charge appearing in the RTA expression:
\begin{align}
    \kappa_B^{\mathrm{RTA}}(T,\mu_B) =
    \frac{1}{3} & \sum_{i=q,\bar q} \int \frac{d^3{\bf p}}{(2\pi)^3} \frac{\mathbf{p}^2}{E_i^2}\tau_i(\mathbf{p},T,\mu_B) \nonumber \\
    & \times \left(b_i - \frac{n_B E_i}{\epsilon+p} \right)^2  d_i f_i \left(1-f_i\right),
    \label{eq:kappa_B_RTA}
\end{align}
where $b_i$ is the baryon number of the parton species, with $b_q=1/3$ for quarks and $b_{\bar q}=-1/3$ for antiquarks, $n_B$ is the net-baryon density, and $\epsilon+p$ is the enthalpy density. Since only quarks and antiquarks carry baryon number, gluons do not contribute directly to $\kappa_B$, although they enter indirectly through the relaxation times.

\begin{figure}[t!]
    \centering
    \includegraphics[width=\columnwidth]{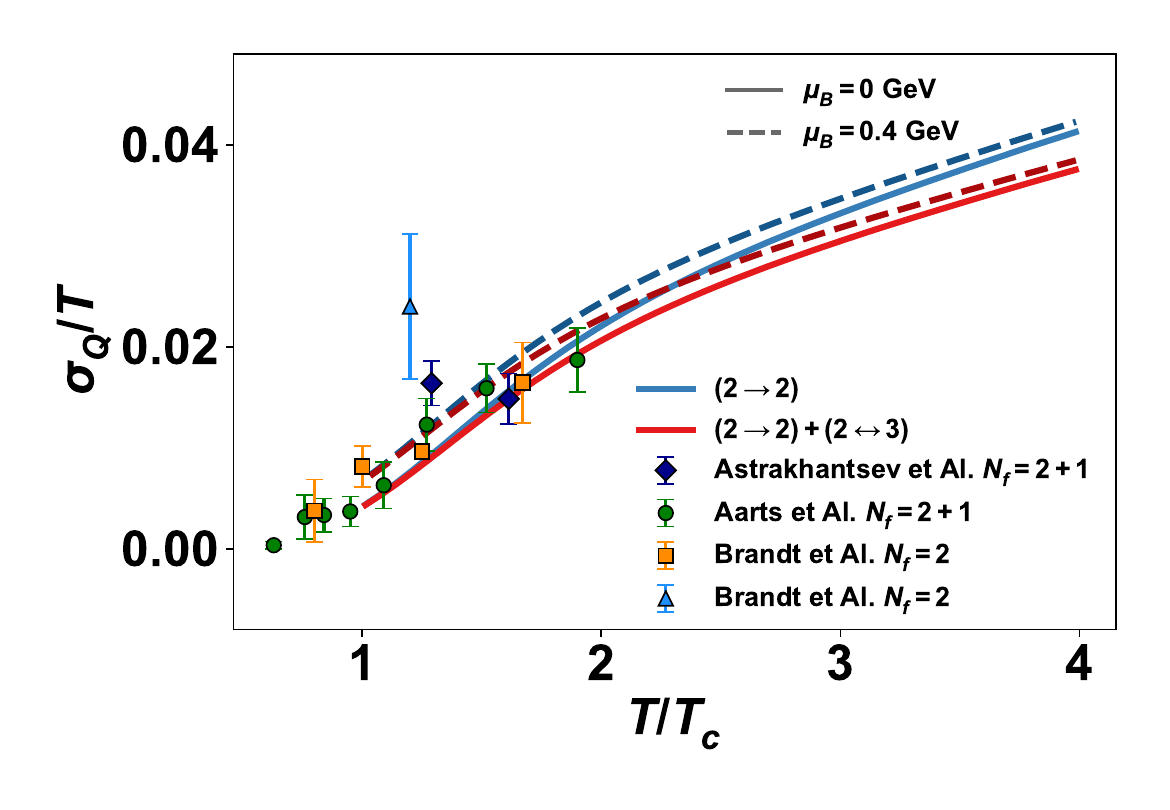}
    \caption{Electric conductivity to temperature ratio $\sigma_Q/T$ as a function of scaled temperature $T/T_C$, at fixed values of baryon chemical potential \GI{$\mu_B=0$ (solid lines), $\mu_B=0.4$ GeV (dashed lines)}. Blue lines show $\sigma_Q/T$ for $2\rightarrow 2$ processes only, while red lines correspond to $\sigma_Q/T$ for both $2\rightarrow 2$ and $2\leftrightarrow3$ processes. Lattice results at $\mu_B=0$ for $N_f=2+1$ from Ref.~\cite{Astrakhantsev:2019zkr} (blue diamonds), Ref.~\cite{Aarts:2014nba} (green circles), and for $N_f=2$ from Ref.~\cite{Brandt:2015aqk} (orange squares), Ref.~\cite{Brandt:2012jc} (cyan triangles) are shown.}
    \label{fig:TC_sigma}
\end{figure}

\begin{figure}[t!]
    \centering
    \includegraphics[width=\columnwidth]{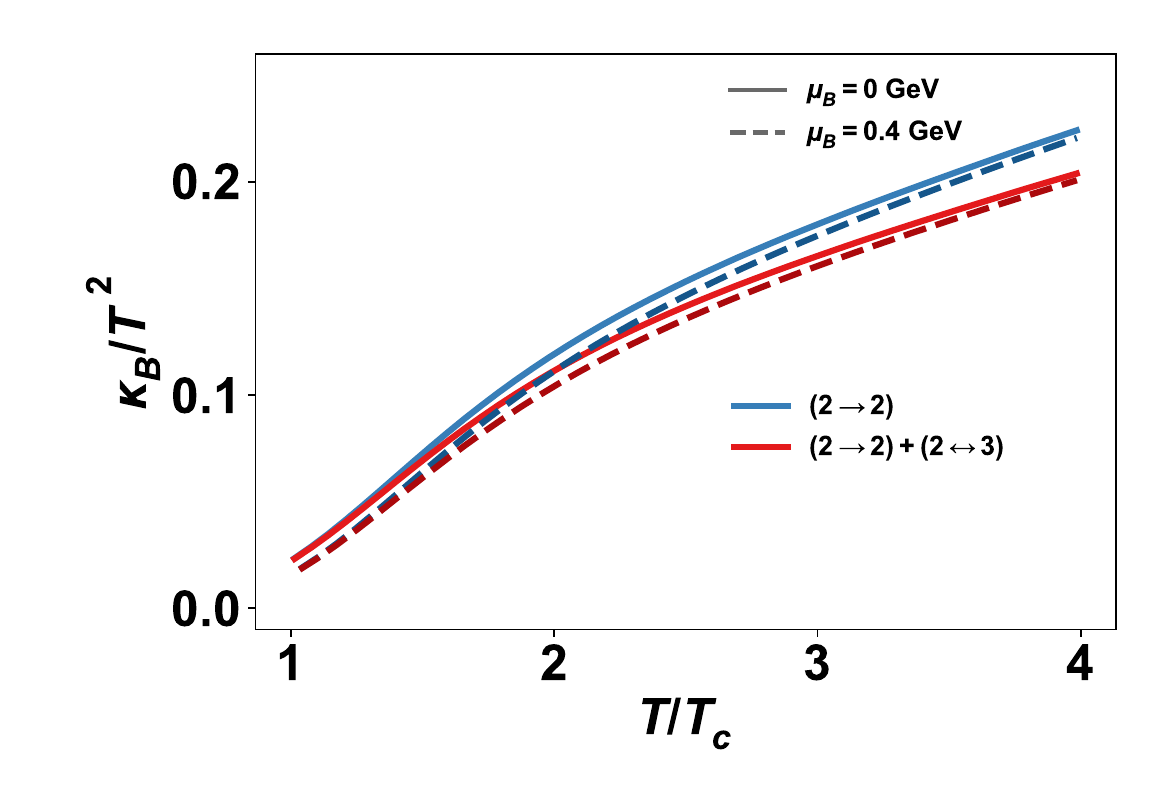}
    \caption{Baryon diffusion to temperature squared ratio $\kappa_B/T^2$ as a function of scaled temperature $T/T_C$, at fixed values of baryon chemical potential \GI{$\mu_B=0$ (solid lines), $\mu_B=0.4$ GeV (dashed lines)}. Blue lines show $\kappa_B/T^2$ for $2\rightarrow 2$ processes only, while red lines correspond to $\kappa_B/T^2$ for both $2\rightarrow 2$ and $2\leftrightarrow3$ processes.}
    \label{fig:TC_kappa}
\end{figure}

The obtained electric conductivity to temperature ratio $\sigma_Q/T$ and baryon diffusion coefficient to temperature squared ratio $\kappa_B/T^2$ are shown as functions of the scaled temperature $T/T_C$ for $\mu_B=0$ and $\mu_B=0.4$ GeV in Figs.~\ref{fig:TC_sigma} and~\ref{fig:TC_kappa}, respectively. Consistent with the behavior observed for the viscosities, the inclusion of gluon-radiation and absorption processes leads to a small reduction of both $\sigma_Q/T$ and $\kappa_B/T^2$, due to the corresponding decrease of the quark and antiquark relaxation times.

The effect remains moderate over the whole temperature range considered, confirming that inelastic processes provide a subleading correction to the elastic contribution also for charge and baryon-number transport. The comparison between $\mu_B=0$ and $\mu_B=0.4$ GeV shows only mild finite-density effects, with no qualitative change in the temperature dependence of either coefficient.


\section{Summary}
\label{sec:summary}

In this work, we have employed the DQPM $2\leftrightarrow3$ scattering channels to assess the impact of inelastic processes on the transport coefficients of baryon-rich quark-gluon plasma, beyond the $2\rightarrow2$ baseline established in previous DQPM studies. To this end, we first quantified the contribution of $2\leftrightarrow3$ reactions to the thermal relaxation times of quarks and gluons. We then evaluated, within the relaxation time approximation, the shear viscosity, bulk viscosity, electric conductivity, and baryon diffusion coefficient as functions of temperature and baryon chemical potential, using microscopic momentum-dependent interaction rates for both \GI{binary and number-changing} scattering processes. 
In this implementation, \GI{the forward gluon-emission $2\rightarrow3$ and inverse gluon-absorption $3\rightarrow2$ channels are included on an equal footing as separate contributions to the quasiparticle relaxation rate within the RTA. Their corresponding interaction rates are found to be comparable. Nevertheless, both contributions remain subleading compared to the dominant $2\rightarrow2$ reactions.}

We found that the inclusion of $2\leftrightarrow3$ channels systematically reduces all four transport coefficients with respect to the $2\rightarrow2$ only baseline, as expected from the corresponding decrease of the relaxation times. In the thermal regime explored here, however, this correction remains moderate, since the inelastic rates stay below the elastic ones over the considered $(T,\mu_B)$ range and become more relevant mainly for parton scatterings with large momenta, which are suppressed in the thermal QGP medium. 
Consequently, the $2\leftrightarrow3$ sector provides a quantitative correction rather than a qualitative modification of the temperature and baryon chemical potential dependence of the transport coefficients.\\
\EB{The moderate impact of the $2\leftrightarrow3$ processes reflects the restricted phase space for massive quasiparticle-gluon emission in the forward reactions and the limited thermal abundance of the additional incoming gluon in the inverse reactions. Within the RTA, these channels increase the total interaction rate and therefore reduce the relaxation time, leading to a slight decrease of the transport coefficients.
}

At $\mu_B=0$, the resulting $\eta/s$, $\zeta/s$, and $\sigma_Q/T$ are compatible with available lattice QCD estimates within uncertainties. At finite $\mu_B$, the results provide model predictions for baryon-rich QCD matter relevant to beam energy scan experiments. Overall, this study supports the robustness of previous DQPM transport coefficient calculations and provides updated microscopic input for hydrodynamic and transport simulations of heavy-ion collisions at finite net-baryon density.


\begin{acknowledgments}
The authors acknowledge inspiring discussions with  W. Cassing, G. Moore and O. Kaczmarek.
 We also acknowledge the support by the Deutsche Forschungsgemeinschaft (DFG) through the grant CRC-TR 211 "Strong-interaction matter under extreme conditions" (Project number 315477589 - TRR 211).  The computational resources utilized for this work were provided by the Center for Scientific Computing (CSC) at Goethe University Frankfurt.
\end{acknowledgments}

\bibliography{references}

\end{document}